\documentstyle[prl,aps,epsf]{revtex}
\def\ltwid{\raise.3ex\hbox{$<$\kern-.75em\lower1ex\hbox{$\sim$}}}

\begin{document}
\twocolumn[\hsize\textwidth\columnwidth\hsize\csname
@twocolumnfalse\endcsname

\draft
\title{
Continuum Limit of Lattice QCD with Staggered Quarks in the Quenched 
Approximation --- A Critical Role for the Chiral Extrapolation
}
\author{
Claude Bernard$^1$, Tom Blum$^2$, Carleton DeTar$^3$,
Steven Gottlieb$^4$, Urs M.~Heller$^5$, James E.~Hetrick$^6$,
Craig McNeile$^3$,
K.~Rummukainen$^7$, R.~Sugar$^8$, D.~Toussaint$^9$
}
\address{
$^1$Washington University, St.~Louis, Missouri 63130, USA\\
$^2$Brookhaven National Laboratory, Upton, New York 11973-5000, USA\\
$^3$University of Utah, Salt Lake City, Utah 84112, USA\\
$^4$Indiana University, Bloomington, Indiana 47405, USA\\
$^5$SCRI, The
Florida State University, Tallahassee,
   Florida 32306-4130, USA\\
$^6$University of the Pacific, Stockton, CA 95211-0197, USA\\
$^7$Nordita, Blegdamsvej 17, DK-2100 Copenhagen \O, Denmark\\ 
$^8$University of California, Santa Barbara, California 93106, USA\\
$^9$University of Arizona, Tucson, Arizona 85721, USA\\
}
\date{\today}

\maketitle

\begin{abstract}\noindent

We calculate the light quark spectrum of lattice QCD in the quenched
approximation using staggered quarks.  
We 
take the light quark mass, infinite volume, continuum limit.
With non-linear chiral extrapolations, we find that the nucleon to $\rho$
mass ratio is $m_N/m_\rho= 1.254\pm 0.018\pm0.028$, where the 
errors are statistical
and systematic (within the quenched approximation), respectively.  
Since the experimental value is 1.22, our results indicate
that the error due to quenching is $\ltwid 5\%$.

\end{abstract}
]

The calculation of the light hadron spectrum is one of the main goals of
lattice QCD.  
However, the computed value of the nucleon to $\rho$ mass ratio has persistently
been too large. 
This has been true for calculations done with
dynamical fermions, or in the quenched approximation
\cite{QUENCHED},  
and for both the Wilson and staggered 
methods of putting quarks on the lattice.\cite{SGREVIEW}.
In such calculations, one must use a finite-volume box and a non-zero lattice
spacing.  In addition, because of the iterative algorithms
used to calculate the quark propagators, 
quark masses larger than those of the
up and down quarks are necessary.  
Thus, before 
comparing
with experiment, one must take the infinite volume, zero lattice 
spacing, and light quark limits to remove systematic errors.
Our extensive series of calculations with 
staggered quarks
in the quenched approximation provides good control over each
of these extrapolations \cite{MILC,MILCNEW}.  
Here we present a new analysis of the chiral extrapolation fully consistent
with quenched chiral perturbation theory \cite{CPT}.

It is important to demonstrate that calculations with
Wilson and staggered quarks have the same continuum limit.
Butler {\it et al}. \cite{GF11} calculated the quenched Wilson
quark spectrum using three lattice spacings.
Their results had a statistical accuracy of 5-6\%, and they
claimed good agreement with the
real world for a number of hadron mass ratios.
However, there are concerns about how well they were able to control
their extrapolations \cite{GF11concern,MILCNEW}.
Recently, the CP-PACS collaboration has presented preliminary results
with Wilson quarks, using
four values of the gauge coupling on large lattices \cite{CPPACS}.



In Table~\ref{kslattices}, 
we list the coupling, volume and size of each or our ensembles of 
lattices.  
(Computational details are found in Ref.~\cite{MILCNEW}.)
%
We used five
quark masses for each ensemble, each a factor of two different from the next
heavier or lighter. 
To increase our statistics,
we calculated propagators from every eighth time slice, so we have four to 
eight sets of hadron propagators per lattice (depending on $N_t$).
These propagators are correlated, so they are blocked
together for further analysis.  

Now let us turn to the sources of systematic errors.
The initial motivation for these calculations was to
understand finite volume effects. \cite{EARLYMILC}.  
We have carried out detailed studies at gauge coupling $6/g^2=5.7$
using lattices with spatial dimensions $N_s=8$, 12, 16, 20, 24, with
the same five quark masses on all five lattices.
At $6/g^2=5.85$, we use three sizes, $N_s=12$, 20 and 24.  
The last column of Table~\ref{kslattices} contains the box size in fm based
on a scale set by the $\rho$ mass extrapolated to the physical quark mass.
In each case, the smallest size is 1.8 fm, considerably smaller than the
2.7 fm size for  $6/g^2=6.15$.
We expect  that in a small volume
the mass will increase as the hadron becomes squeezed by the box, and that
this effect increases for smaller quark masses.
For the $\rho$ there is very little evidence for finite size effects.  
Comparing the smallest and largest sizes for the three lightest quark masses,
the largest difference is $2.5 \pm 1.3 \%$ and the $\rho$ is lighter on the
smaller size.
We expect somewhat larger finite size effects for the nucleon.  For
$6/g^2=5.85$ we find a $3.2 \pm 1.1 \%$ effect for the lightest quark
mass; however, at 5.7 for the corresponding quark mass the effect is only
$1.0\pm1.2 \%$.
For small box sizes, there is evidence that
finite-size effects fall like $1/V$ \cite{FS1}, where
$V$ is the spatial lattice volume.  For larger boxes, an exponential decay
proportional to $\exp(-m_\pi L)$ is expected \cite{FS2}.  By the former
consideration, the finite size effect
for our 2.7 fm $6/g^2=6.15$ lattice should be smaller by 
a factor of $(2.7/1.8)^3 \approx 3.4$.
By the latter, there should be a decrease by a factor of $\approx\! 5$.
Thus, we expect finite size effects smaller than 1\% for $6/g^2=6.15$,
and even smaller effects for the two
largest volumes at $6/g^2=5.7$ and 5.85.

The chiral extrapolation requires the most care and controls
the final error.  
It is based on a small mass expansion,  but since we cannot generate accurate
results for small quark masses, we are forced to use intermediate and
heavy quark masses.  In this mass region the contribution of the
higher order terms in the chiral expansion can be significant, 
and the extrapolation to the physical light quark masses has to
be performed very carefully.
Another complication arises from 
the quenched approximation.
Quenched chiral perturbation theory ($Q\chi PT$) \cite{CPT}
differs from the ordinary chiral perturbation theory ($\chi PT$)
that applies to the real world. 

Our lightest three masses cover a range of a factor of four, and
cannot be fit to a linear function with even a marginally acceptable confidence level (CL).  
(The chiral fits are done using the full covariance matrix of the hadron masses
for different quark masses.)
We began by trying a dozen different fitting forms
(Table~\ref{fittingforms}).
The term proportional to
 $m_q$ appears at tree-level (in
both $\chi PT$ and $Q\chi PT$) and is therefore expected to
be the most important correction to the chiral limit for a ``reasonable''
range of masses. The other $m_q$ dependent terms should be
thought of as one-loop corrections: $m_q^{3/2}$, $m_q^2$, and $m_q^2 \log m_q$
are standard terms which appear in both
$\chi PT$ and $Q\chi PT$; while
the more singular (at $m_q=0$)
terms proportional to $\sqrt{m_q}$ and $ m_q \log m_q$
arise only in  $Q\chi PT$, and are due to $\eta'$ loops \cite{QCHPTNUCRHO}.
However, we reject
fits 1,2,3,4 and 10, which include an $\sqrt{m_q}$ term,
because the
coefficient of $\sqrt{m_q}$ found by the fits
is opposite in sign to what is expected from $Q\chi PT$.
In Fig.~1, we show seven fits for the nucleon at $6/g^2=6.15$ that have
good CL.  As can be seen in the inset, several fits
(2, 3, 4, 10) decrease sharply at small quark mass.  However, a
$\sqrt{m_q}$ term with the sign implied by $Q\chi PT$ would cause
a sharp increase.

In addition, due to flavor breaking,
the $\sqrt{m_q}$ term is not really appropriate to quenched staggered
quarks.
It is actually the flavor singlet pion that appears \cite{SHARPE}
in $Q\chi PT$, and its
mass is not proportional to $\sqrt{m_q}$ at finite lattice spacing.
Thus, we should use instead a term proportional to
the mass of the non-Goldstone pion, commonly denoted $m_{\pi_2}$.
Since $Q\chi PT$ gives us only a rough
idea of the values of the proportionality constants, we consider a range
of values. 
In practice, we define, for fixed $\lambda_1$ and $\lambda_2$,
\begin{eqnarray}
m'_N \equiv (m_N + \lambda_1 m_{\pi_2}){m_N^{\rm phys}
\over m_N^{\rm phys} +  \lambda_1 m_{\pi}^{\rm phys}} \\
m'_\rho \equiv (m_\rho + \lambda_2 m_{\pi_2}){m_\rho^{\rm phys}\over
m_\rho^{\rm phys} +  \lambda_2 m_{\pi}^{\rm phys}},
\label{primedmasses}
\end{eqnarray}
where ``phys'' stands for the physical values and the other quantities
are values computed at given quark mass and lattice spacing.
We then fit $m'_N$ and $m'_\rho$ 
to functions 8 and 12 (Table~\ref{fittingforms})
for  various values of $\lambda_1$ and $\lambda_2$ obeying
$0.0\le\lambda_1 \le \lambda_2\le 0.4$, which is the expected
range of values from   $Q\chi PT$ \cite{QCHPTNUCRHO,CHPTNUC}.

Table~\ref{confidence} contains the combined CL of these chiral
fits for our five lattices with the weakest couplings and largest
volumes:
$N_s=32$ at $6/g^2=6.15$, and
$N_s=20$ and 24 at $6/g^2=5.7$ and
5.85.  
Although many of these fits
are marginally acceptable, none of the CL are very good. 
This may be due to the fact that $Q\chi PT$ to order $m_q^2$ would require all the terms that appear in
either Fit 8 or 12; while
we are limited 
to a maximum of 4 fit parameters since we have only 5 masses at our disposal.
It is nevertheless encouraging that almost
all the fit parameters are of the
rough size (within a factor of 2)
and sign predicted by $Q\chi PT$ \cite{QCHPTNUCRHO,CHPTNUC}.
The exception is the
$m_q^{3/2}$ term in the $\rho$ fits: its coefficient is more than
an order of magnitude smaller than the size suggested in \cite{QCHPTNUCRHO}.
Such agreement is as good as could be expected, since the $Q\chi PT$
predictions are currently based either on rather arbitrary
guesses of parameters ({\it e.g.}\ the
$m_q^{3/2}$ case for the $\rho$) or on parameter estimates
taken from real-world ({\it i.e.}, not quenched)
data and determined only up to large errors.
As expected, the linear term in $m_q$
is dominant; all other terms give small corrections except
at the largest values of the quark mass. The fit parameters are relatively
stable: As the fit (8 or 12) or $\lambda$ values are changed,
the parameters change by at most $\pm25\%$ from their average
values, while the masses extrapolated to the chiral limit
are considerably more stable (see Table \ref{STEVESNEWTABLE}).
The range of results for the acceptable fits gives
us an estimate of the chiral extrapolation error.
All the other fit functions 
in Table~\ref{fittingforms} are
either inconsistent with $Q\chi PT$ or have very low CL:
the highest of these is $\sim0.0001$.

The final extrapolation is in the lattice spacing $a$.  
Although we expect $O(a^2)$ errors in general for the staggered action,
flavor symmetry breaking gives $m_{\pi_2}^2 \approx A a^2$ at
$m_q=0$ \cite{PITWO}, which implies that $m_{\pi_2}$ contributions to the
nucleon and $\rho$ could produce $O(a)$ terms in $m_N/m_\rho$.  
Since, however, we
are attempting to {\em remove} the $m_{\pi_2}$ dependence, we believe
it is still reasonable to fit to the quadratic form
$C + B (a m_\rho)^2$ for our central values.
In Fig.~\ref{ratio0808vsa}, 
we show the lattice spacing extrapolation 
based on chiral fit 8 with $\lambda_1 = 0.0$ for the nucleon and 
fit 12 with $\lambda_2 = 0.1$ for the $\rho$, which give the highest CL.
For the $\pi$ chiral extrapolation, which is needed to determine
the physical light quark mass,
we use the form $m_\pi^2=am_q + bm_q^2+cm_q^3+ d m_q\ln m_q$.  
When we keep only the 
three weakest couplings,
the continuum extrapolation with these parameters
has a good CL (0.74) and gives a result which is 
close to the overall average of the continuum extrapolations of 
all the chiral fits.  For each of the two intermediate couplings, we
plot the two largest volumes, but only the largest volume was included
in the fit to the $a$ dependence.  
Extrapolating to the continuum limit we get 1.251 $\pm$ 0.018.

To obtain our central value, we look at 
all (quadratic) continuum extrapolations 
of the three weakest couplings coming from
any of the chiral fits in Table~\ref{confidence} with CL 
greater than 0.04.  (The continuum
extrapolations 
all have ${\rm CL}>0.34$.)
Figure \ref{ratio_vs_confidence} shows the results of these extrapolations
of $m'_N/m'_\rho$ as a function of the CL of the
continuum extrapolation.
After averaging over all fits in Fig.~\ref{ratio_vs_confidence},
we obtain $m_N/m_\rho= 1.254\pm 0.018\pm0.022$, where the last
error is the standard deviation over the fits, which we take
as a combination of chiral extrapolation and continuum extrapolation errors.  
The result is rather insensitive to the cut on
chiral CL: changing  the cut to 0.005 or 0.06 
(a cut greater than 0.06 would
rule out all the $\rho$ fits in Table~\ref{confidence}) 
changes the central value by 0.010 or 0.008, respectively.

To explore further the error in the continuum extrapolation, we then include
the strongest coupling point and repeat the analysis.  
Averaging  over all fits with chiral and continuum CL
greater than 0.04 gives $1.248\pm0.016\pm0.008$. 
We then change the continuum extrapolation to include
the higher power $(am_\rho)^4$, as well as $(am_\rho)^2$.  
By averaging as before, we obtain $1.266\pm0.020\pm 0.021 $. 

Because the true values of $\lambda_{1,2}$ are not known, we
cannot definitively subtract off the $m_{\pi_2}$ dependence. Therefore,
there may remain a small $O(a)$ term in $m'_N/m'_\rho$.  To study this
effect, we  add a linear term in $a$ to our central continuum
extrapolation fits above.  (These are now constrained fits.)
The coefficient of the linear term is always consistent
with 0, with the errors giving a bound on its magnitude of the size
expected from a small residual $m_{\pi_2}$ contribution.  Averaging
the extrapolated values of $m'_N/m'_\rho$ gives $1.269\pm0.096\pm 0.016 $.

These considerations lead us to include additional errors of
0.015 (the effect changing the continuum extrapolation) and
0.010 (the effect of changing the cutoff on the CL). 
Combining them in quadrature with the 0.022 determined above,
gives a total systematic error due to chiral and continuum extrapolations
of  0.028.

In summary, 
the chiral extrapolation is the most delicate issue in our
computation of the nucleon to the $\rho$ mass ratio.
The simple linear
chiral extrapolation is ruled out.  
Our results are reasonably well described by fits motivated by $Q \chi PT$.
However, the CL of such fits is a slowly varying function of 
the parameters $\lambda_1$ and $\lambda_2$, which therefore are not determined
in our procedure.  Instead, 
we average over reasonable ranges of these
parameters. Fortunately, the continuum values of $m_N/m_\rho$
produced are not strongly dependent on the 
parameters.  Taking into account the variance over the fits, we obtain
$m_N/m_\rho= 1.254\pm 0.018\pm0.028$.
Comparison with the observed value of 1.22 indicates that the effects
of quenching are less than about 5\% at the 1 $\sigma$ level.

This work was supported by the U.S. DOE and NSF.
Calculations were done at Indiana University,
Pittsburgh Supercomputing Center, NERSC and Sandia NL.
We also thank  
M.~Golterman,
S.~Sharpe and A.~Ukawa.

\begin{figure}[thb]
\epsfxsize=0.8 \hsize
\epsffile{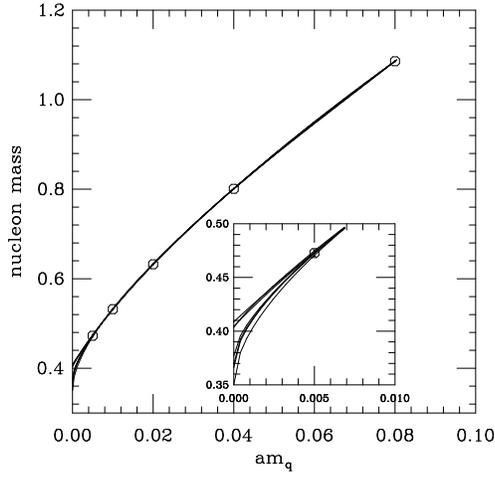}
\caption{The nucleon mass {\it vs}.\ quark mass with $6/g^2=6.15$.  
We plot fits 12, 8, 9, 10, 3, 4 and 2 (listed in order of decreasing
chiral limit), all of which have CL greater than 0.45.}
\label{nuc615}
\end{figure}

\begin{figure}[thb]
\epsfxsize=0.8 \hsize
\epsffile{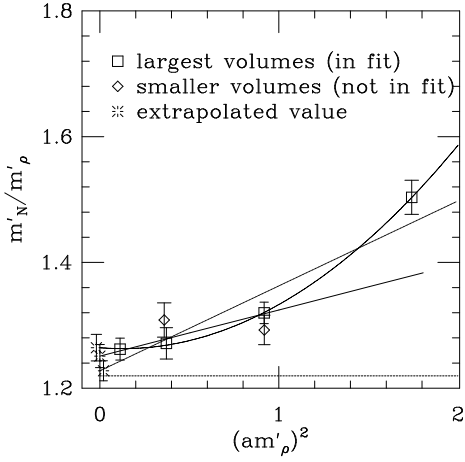}
\caption{$m_N/m_\rho$ at the physical quark mass {\it vs}.\ $(am_\rho)^2$.
The points come from fit 8 for the nucleon and 12 for the $\rho$,
with $\lambda_1 = 0.0$, $\lambda_2=0.1$.
The horizontal line indicates the physical value.  
The solid straight line shows a linear (in $(am_\rho)^2$) extrapolation that includes only the
three smallest lattice spacings;  
the dot-dashed straight line, all four
lattice spacings.
The curve is a higher order fit to all four
spacings.
The extrapolated 
values near $am_\rho=0$ have been spread horizontally for clarity.}
\label{ratio0808vsa}
\end{figure}

\begin{figure}[thb]
\epsfxsize=0.8 \hsize
\epsffile{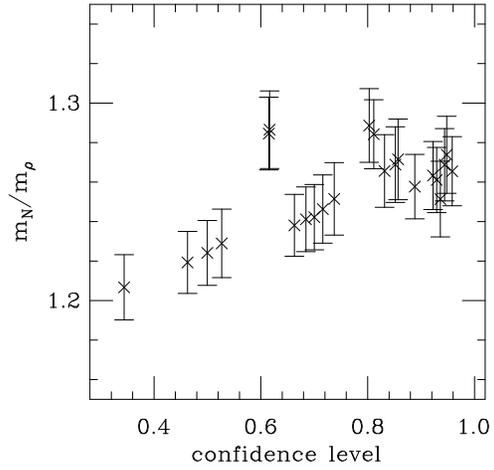}
\caption{$m'_N/m'_\rho$ extrapolated to $a=0$ {\it vs}.\ the 
CL of the fit used for this extrapolation. 
The points represent different choices for the
chiral fits of the $\rho$ and nucleon and different values of the
parameters $\lambda_1$ and $\lambda_2$.
All extrapolations in $a$ are quadratic and keep only the three
weakest couplings.
The combined CL over these couplings of each chiral fit included here is
greater than 0.04.
}
\label{ratio_vs_confidence}
\end{figure}

\begin{table}[tb]
\caption{Lattices used for spectrum calculation}
\label{kslattices}
\begin{tabular}{lcccc}
$6/g^2$&size&\#&$am_q$&$aN_s$ (fm)\\
$ 5.54 $&   $16^3 \times 32$&          205 &0.02--0.32&4.96\cr
\noalign{\smallskip}
 $ 5.7 $&   $8^3 \times 48$&          605 & 0.01--0.16&1.83\cr
 $ 5.7 $&   $12^3 \times 48$&          405 & 0.01--0.16&2.75\cr
 $ 5.7 $&   $16^3 \times 48$&          405 & 0.01--0.16&3.66\cr
 $ 5.7 $&   $20^3 \times 48$&          205 & 0.01--0.16&4.58\cr
 $ 5.7 $&   $24^3 \times 48$&          199 & 0.01--0.16&5.49\cr
\noalign{\smallskip}
 $ 5.85 $&   $12^3 \times 48$&          205 & 0.01--0.16&1.76\cr
 $ 5.85 $&   $20^3 \times 48$&          205 & 0.01--0.16&2.93\cr
 $ 5.85 $&   $24^3 \times 48$&          200 & 0.01--0.16&3.52\cr
\noalign{\smallskip}
 $ 6.15 $&   $32^3 \times 64$&          115 & 0.005--0.08&2.65\cr
\end{tabular}
\end{table}

\begin{table}
\caption{Our fitting functions}
\label{fittingforms}
\begin{tabular}{ll}
\rm Fit 1:&$M+ a m_q^{1/2}$\\
\rm Fit 2:&$M+ a m_q^{1/2} + b m_q$\\
\rm Fit 3:&$M+ a m_q^{1/2} + b m_q +c m_q^{3/2}$\\
\rm Fit 4:&$M+ a m_q^{1/2} + b m_q +c m_q^2$\\
\rm Fit 5:&$M+ a m_q$\\
\rm Fit 6:&$M+ a m_q  +b m_q^{3/2}$\\
\rm Fit 7:&$M+ a m_q  +b m_q^2$\\
\rm Fit 8:&$M+ a m_q  +b m_q^{3/2} + c m_q^2$\\
\rm Fit 9:&$M+ a m_q  +b m_q \log m_q$\\
\rm Fit 10:&$M+ a m_q^{1/2} + b m_q +c m_q\log m_q$ \\
\rm Fit 11:&$M+ a m_q  +b m_q^2 \log m_q$\\
\rm Fit 12:&$M+ a m_q  +b m_q^2 + c m_q^2 \log m_q$\\
\end{tabular}
\end{table}

\begin{table}[tb]
\caption{Combined CL of chiral fits}
\label{confidence}
\begin{tabular}{llllll}
Fit&$\lambda=0.0$&$\lambda=0.1$&$\lambda=0.2$&$\lambda=0.3$&$\lambda=0.4$\\
\hline
\multicolumn{6}{c}{Nucleon Jackknife fits}\\
\hline

8 & 0.186 & 0.140 & 0.103 & 0.075 & 0.054 \\
12 & 0.086 & 0.050 & 0.028 & 0.015 & 0.008 \\
\hline
\multicolumn{6}{c}{Rho Jackknife fits}\\
\hline
8 &0.048& 0.041 & 0.035 & 0.031 & 0.028 \\
12 &0.059& 0.060 & 0.054 & 0.046 & 0.037 \\
\end{tabular}
\end{table}

\begin{table}[tb]
\caption{Dependence of nucleon and $\rho$ masses on chiral fit
at $6/g^2=6.15$. $\tilde m_N\equiv m_N+\lambda_1 m_{\pi_2}$
extrapolated to $m_q=0$. ${\rm ``}m_N{\rm "}\equiv \tilde m_N -
\lambda_1 m_{\pi_2}$, where $m_{\pi_2}$ is separately extrapolated
to $m_q=0$ with the same fit.  Similarly for $\rho$.}
\label{STEVESNEWTABLE}
\begin{tabular}{lllll}
&$\!\!\!\!$Fit&$\!\lambda_1=0.0$&$\!\lambda_1=0.2$&$\!\lambda_1=0.4$\\
\hline

$\tilde m_N$& $\!\!\!\!$8 & 0.404(6) & 0.427(5) & 0.450(5) \\
${\rm ``}m_N{\rm "}$& $\!\!\!\!$8 & 0.404(6) & 0.407(5) & 0.410(5) \\
$\tilde m_N$& $\!\!\!\!$12 & 0.409(5) & 0.433(5) & 0.457(5) \\
${\rm ``}m_N{\rm "}$& $\!\!\!\!$12 & 0.409(5) & 0.413(5) & 0.417(5) \\
$\tilde m_\rho$& $\!\!\!\!$8 & 0.320(4) & 0.343(4) & 0.366(4) \\
${\rm ``}m_\rho{\rm "}$& $\!\!\!\!$8 & 0.320(4) & 0.323(4) & 0.326(4) \\
$\tilde m_\rho$& $\!\!\!\!$12 & 0.319(3) & 0.343(3) & 0.367(3) \\
${\rm ``}m_\rho{\rm "}$& $\!\!\!\!$12 & 0.319(3) & 0.323(3) & 0.327(3) \\
\end{tabular}
\end{table}


\begin{references}
\bibitem{QUENCHED}
E. Marinari, G.~Pa\-ri\-si and C.~Rebbi, Nucl. Phys. {\bf B190}, 734 (1981);
H.~Hamber and G.~Parisi, Phys. Rev. Lett. {\bf 47}, 1792 (1981);
D.~Weingarten, Phys. Lett. {\bf 109B}, 57 (1982).

\bibitem{SGREVIEW} 
S.~Gottlieb, 
Nucl. Phys. B (Proc. Suppl.) {\bf 53}, 155 (1997),
contains a more detailed discussion of the finite volume and
chiral extrapolations with both Wilson and staggered quarks.

\bibitem{MILC} 
Some preliminary results have been presented in C.~Bernard {\it et al}.,
Nucl. Phys.~B, (Proc. Suppl.), {\bf47}, 345 (1996); 
{\bf53}, 212 (1997).

\bibitem{MILCNEW}
C.~Bernard {\it et al}.,
Nucl. Phys.~B, (Proc. Suppl.), {\bf60A}, 3 (1998).

\bibitem{CPT}
S.~R.~Sharpe, Phys. Rev. D {\bf41}, 3233 (1990); 
{\bf46}, 3146 (1992); 
C.~Bernard and M.~Golterman, Phys. Rev. D {\bf46}, 853 (1992); 
Nucl. Phys.~B (Proc. Suppl.) {\bf26}, 360 (1992).

\bibitem{GF11}
F.~Butler, H.~Chen, J.~Sexton, A.~Vaccarino and D.~Weingarten,
Nucl. Phys. B {\bf 430}, 179 (1994); Phys. Rev. Lett. {\bf70}, 2849
(1993).

\bibitem{GF11concern}
T.~Bhattacharya, R.~Gupta, G.~Kilcup and S.~Sharpe, Phys. Rev. D{\bf 53},
6486 (1996).

\bibitem{CPPACS}
S.~Aoki {\it et al}., 
Nucl. Phys.~B, (Proc. Suppl.), {\bf60A}, 14 (1998);
Report No. hep-lat/9709139.


\bibitem{EARLYMILC}
S.~Gottlieb, Nucl. Phys.~B. (Proc. Suppl.), {\bf42}, 346 (1995).


\bibitem{FS1}
M. Fukugita {\it et al}., Phys. Lett. B {\bf 294}, 380 (1992).

\bibitem{FS2}
M.~L\"uscher, Nucl. Phys. {\bf B354}, 531 (1991).

\bibitem{QCHPTNUCRHO}
J.~N.~Labrenz and S.~R.~Sharpe, Nucl. Phys. B (Proc. Suppl.) 
{\bf34}, 335 (1994);
Phys. Rev. D {\bf54}, 4595 (1996). M. Booth, G. Chiladze, and A. Falk,
Phys. Rev. D {\bf 55}, 3092 (1997). 

\bibitem{SHARPE} S.~R.~Sharpe, Nucl. Phys. (Proc. Suppl.) {\bf53}, 181 (1997).

\bibitem{CHPTNUC} 
E.\ Jenkins and A.V.\ Manohar, Phys.\ Lett.\ B {\bf 255}, 558 (1991);
{\bf 259}, 353 (1991); 
V.\ Bernard {\it et al.}, Z.\ Phys.\ {\bf C60}, 11 (1993).
 
\bibitem{PITWO} S.~R.~Sharpe {\it et al.},  Nucl. Phys. B (Proc. Suppl.)
{\bf 26}, 197 (1992);
S.\ Aoki {\it et al.}, 
Nucl. Phys. B (Proc. Suppl.) {\bf 53}, 209 (1997).
 From our data, we find $A\approx m_\rho^4$, consistent with
S.\ Aoki {\it et al.}

\end{references}
\end{document}